\documentstyle[aps,pre,epsf,floats]{revtex}
\begin{document}
\draft
\twocolumn[\hsize\textwidth\columnwidth\hsize\csname
@twocolumnfalse\endcsname

\title{Surface states in nearly modulated systems\\}

\author{A. E. Jacobs$^{(1,2)}$, D. Mukamel$^{(2)}$ and D. W. Allender$^{(3)}$}

\address{$^{1}$ Department of Physics, University of Toronto, Toronto,
Ontario, Canada M5S 1A7 \\ $^{2}$
         Department of Physics of Complex Systems, The Weizmann Institute
         of Science, Rehovot 76100, Israel\\ $^{3}$
      Department of Physics and Liquid Crystal Institute, Kent State
      University, Kent, Ohio 44242 USA \\ 
      [-2mm] $ $}

\date{\today}

\maketitle
\begin{abstract}

\hskip 0.3in    A Landau model is used to study the phase behavior of
the surface layer for magnetic and cholesteric liquid crystal
systems that are at or near a Lifshitz point marking the boundary
between modulated and homogeneous bulk phases.  The model
incorporates surface and bulk fields and includes a term in the
free energy proportional to the square of the second derivative of
the order parameter in addition to the usual term involving the
square of the first derivative.  In the limit of vanishing bulk
field, three distinct types of surface ordering are possible:  a
wetting layer, a non-wet layer having a small deviation from bulk
order, and a different non-wet layer with a large deviation from
bulk order which decays non-monotonically as distance from the
wall increases. In particular the large deviation non-wet layer is
a feature of systems at the Lifshitz point and also those having
only homogeneous bulk phases.
\end{abstract}
\vspace{2mm}
\pacs{PACS numbers: 61.30.Cz; 64.60.Kw; 64.70.Md}]


\section{Introduction}

The interaction of a bulk system with a wall may give rise to a
large variety surface phenomena, associated with the thermodynamic
behavior of the surface layer adjacent to the wall. For example in
ferromagnetic systems, when the interaction with the wall is such
that it enhances local order it may happen that a surface
transition takes place at temperatures above the critical
temperature of the bulk. In such a transition the layers close to
the wall become ordered although the bulk remains disordered.
Depending on the nature of the interactions within the bulk and
the interactions between the bulk and the wall, the system may
exhibit phenomena such as wetting, critical wetting, prewetting
and other surface phase transitions. These phenomena have been
extensively studied, both theoretically and experimentally in
recent years (for a review see \cite{Dietrich}).

A study of the global phase diagram for surface critical phenomena
in ferromagnetic and other homogeneously ordered systems has been
carried out by Nakanishi and Fisher \cite{Nakanishi}. In this
study, a Landau phenomenological approach has been applied and the
phase diagram has been analyzed in the space of temperature,
surface enhanced interactions, and bulk and surface ordering
fields. For example it has been found that for finite positive
surface field and no surface enhanced interactions, and in the
limit of vanishingly small negative bulk field, the system
exhibits a wetting transition as the temperature is varied below
the bulk ordering temperature. At low temperatures the surface
field induces a local order in a layer of finite thickness $\it l$
near the wall. However, at temperatures just below the bulk
ordering temperature, the thickness surface layer is infinite,
yielding a $"$wet$"$ state. The two regimes are separated by a
first order transition in which the thickness of the layer
undergoes a discontinuous jump. This is known as the wetting
transition.

More recently surface phenomena in {\it modulated systems} have
been considered. These systems are characterized by a periodic
spatial variation of the order parameter in the bulk. Examples are
magnetic spirals, cholesteric liquid crystals, amphiphilic
systems, diblock copolymers and many others. In many cases the
modulated phase is driven by a gradient-squared term with negative
coefficient in the Landau free energy. The system is then
stabilized by terms quadratic in the second derivative. Studies of
surface phenomena in such systems suggest that surface phase
diagrams are rather rich, exhibiting novel surface states
\cite{JMA} and complicated surface structures
\cite{Fredrickson,Andelman1,Andelman2}. However the possible
global phase diagrams of these systems have not been fully
explored.

Systems exhibiting a Lifshitz point may be considered as
intermediate between ferromagnetic and modulated
\cite{Hornreich,Shapira81,Becerra96}. In the Landau free energy of
such systems, the coefficient of the gradient-squared term
vanishes, making the quadratic term in the second derivatives the
leading order interaction term. Surface phase diagrams of these
systems have not been explored so far and it would be of interest
to study them in some detail.

In this paper we study the surface states and the surface phase
diagram corresponding to a model of a Lifshitz point within the
Landau approach. The phase diagram is studied in the space of
temperature, bulk and surface fields. It is found that unlike the
ferromagnetic case, these systems do not exhibit a wet phase in
which the thickness of the surface layer diverges. It rather
exhibits a transition from one surface state to another, as the
temperature is varied, where $\it both$ surface states have a
finite thickness.

We also consider the surface phase diagram of a ferromagnetic
system which is characterized by higher order interaction terms.
Specifically we consider a Landau free energy which includes terms
quadratic in the gradient and in the second derivatives of the
order parameter, both with positive coefficient. As is well known,
the quadratic term in the second derivatives does not affect the
bulk phase diagram as long as the sign of its coefficient is
positive. However we find, rather surprisingly, that although this
term does not introduce any competing with the gradient squared
term, it affects the surface diagram in a profound way. In
particular, we find that in addition to the usual wet and non-wet
states which exist in the model of Nakanishi and Fisher, the model
exhibits a second non-wet state with a distinct structure of the
order parameter near the surface. Numerical studies yield the
global phase diagram of the model.

The paper is organized as follows: in Section II we review the
model of Nakanishi and Fisher, and present analytic expressions
for the location of the wetting and the critical prewetting
points. Results of a numerical study of the surface phase diagram
corresponding to the model with a Lifshitz point are presented in
Section III. The surface states and the surface phase diagram of
the generalized ferromagnetic model are discussed in Section IV.
Finally a short summary is given in Section V.

          %

\section{Ferromagnetic Model}

In this section we consider the surface phase diagram of  a
ferromagnetic system in the space of temperature, and bulk and
surface ordering fields. The phase diagram exhibits wetting,
prewetting and critical prewetting transitions. The Landau
phenomenological model of Nakanishi and Fisher is reviewed, and
analytic expressions for the wetting and the critical prewetting
points are given. In this model the ferromagnetic interaction is
simply introduced by a gradient-squared term with positive
coefficient. Extensions of this model to other ferromagnetic
systems with higher order ferromagnetic interactions, and to
systems exhibiting Lifshitz points will be discussed in the
following sections.

Let $\phi(x)$ be the scalar order parameter which corresponds to
the ferromagnetic order. The wall is taken to be in the $y-z$
plane, and the order parameter is assumed to depend only on the
coordinate $x$ perpendicular to the wall. The Landau free energy
is given by

\begin{equation}
\label{freenergy}
 F=\int_0^L dx(-h \phi
+ {\textstyle{1\over2}}r \phi^2 + {\textstyle{1\over4}}  \phi^4 +
{\textstyle{1\over2}} (\phi^{\prime})^2 ) -h_s \phi_s
\end{equation}
where $\phi_s = \phi(0) $ denotes the value of the order parameter
at the wall and $\phi^{\prime}=d\phi/dx$. The system is assumed to
be of length $L$ in the $x$ direction. In calculating the surface
free energy we will take the limit $L \to \infty$. We have scaled
the order parameter, the energy and the unit of length to simplify
the coefficients, and so the bulk field $h$, the temperature $r$,
and the surface field $h_s$ are rescaled variables.

The $(r,h)$ phase diagram of this model for non-vanishing surface
field $h_s > 0$ is given schematically in Figure 1
\cite{Nakanishi}. The order parameter away from the wall is not
affected by the surface field. For a negative bulk field, $h < 0$,
it approaches a negative value characteristic of the bulk. On the
other hand the surface field enhances the order within a layer of
thickness $\it l$. The thickness of this layer undergoes a
discontinuous change along the prewetting transition line in the
Figure. For $h=0$ this becomes the first order wetting transition,
$WT$. The line ends at some critical point $CP$ known as the
critical prewetting point. The low temperature surface state,
existing to the left of the line is the prewet state, $PW$.
It is characterized by a surface layer with a finite width. The
state to the right of the line is the wet state, $W$, in which the
width of the surface layer diverges in the limit of vanishing
bulk field. In the following we analyze the Landau
free energy (\ref{freenergy}) and obtain analytic expressions for
the wetting $WT$ and the critical prewetting $CP$ points.
%
%
%
%
\begin{figure}
\epsfxsize=7.0cm \centerline{\epsffile{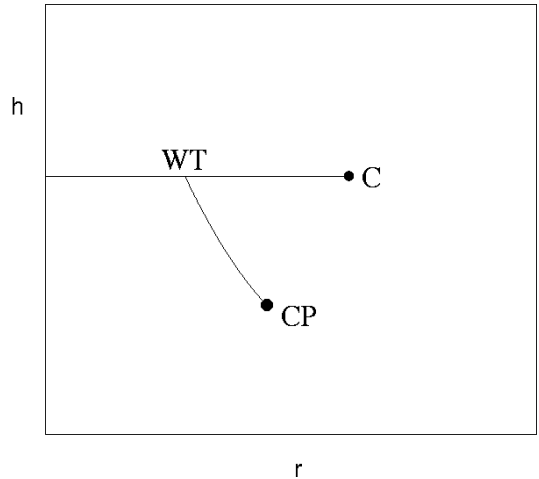}}
\vspace{0.3cm} \caption{A schematic $(r,h)$ phase diagram of the
Nakanishi-Fisher model for $h_s > 0$. The horizontal line
represents the bulk coexistence line, which terminates at an
ordinary critical point $C$. The curved line at negative bulk fields
is the prewetting transition line on which the two types of
surface solutions coexist. This line intersects the $h=0$ axis at
the wetting point $WT$ which separates the dry phase existing at
large negative $r$, known as the prewet phase,$PW$ from the wet phase,
$W$, which exists close to the
critical point. The prewetting transition line terminates at the critical
prewetting point CP, where the two surface solutions become
identical. The width $\it l$ of the surface layer of the wet and
the prewet phases is finite for non-vanishing bulk field. However $\it
l$ diverges in the limit $h \to 0^-$ in the wet phase.}
\end{figure}
The Euler-Lagrange equation corresponding to the free energy
\ref{freenergy} is
\begin{equation}
\label{euler} \phi^{\prime\prime} +h - r\phi-\phi^3=0\
\end{equation}
with the boundary condition at $x=0$
\begin{equation}
\label{boundary} h_s = -\phi_s^{\prime}
\end{equation}
We are interested in calculating the order parameter profile and
free energy for negative bulk field $h<0$ and positive surface
field $h_s > 0$. We thus expect that for large $x$, the order
parameter approaches the bulk value $- \phi_B$ where $\phi_B > 0$
satisfies
\begin{equation}
\label{bulkOP} h + r \phi_B + \phi_B^3 = 0
\end{equation}
Multiplying Eq. (\ref{euler}) by $\phi^\prime$ this equation may
be integrated to yield
\begin{equation}
\label{integratedEQ} {\textstyle{1\over2}}r \phi^2 +
{\textstyle{1\over4}}  \phi^4 -h \phi - {\textstyle{1\over2}}
(\phi^{\prime})^2  =C
\end{equation}
where $C$ is a constant. This constant may be evaluated by noting
that at large $x$ the order parameter asymptotically approaches
$-\phi_B$. Thus
\begin{equation}
\label{C} C={\textstyle{1\over2}}r \phi_B^2 +
{\textstyle{1\over4}}  \phi_B^4 +h \phi_B
\end{equation}
Using this result, the first integral of the Euler equation
(\ref{euler}) becomes
\begin{figure}
\epsfxsize=7.0cm \centerline {\epsffile{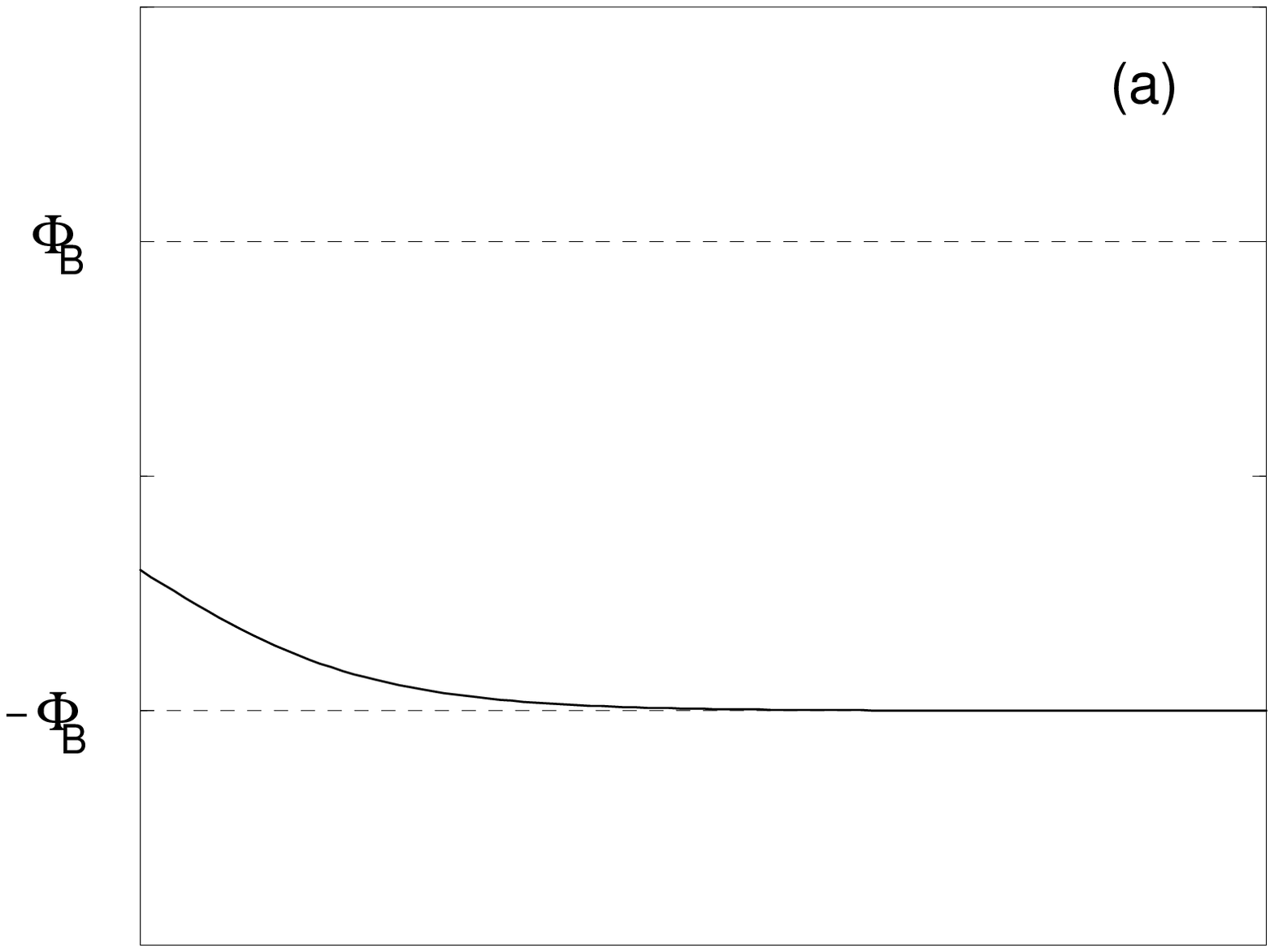}}
\vspace{0.5cm} \epsfxsize=7.0cm
\centerline{\epsffile{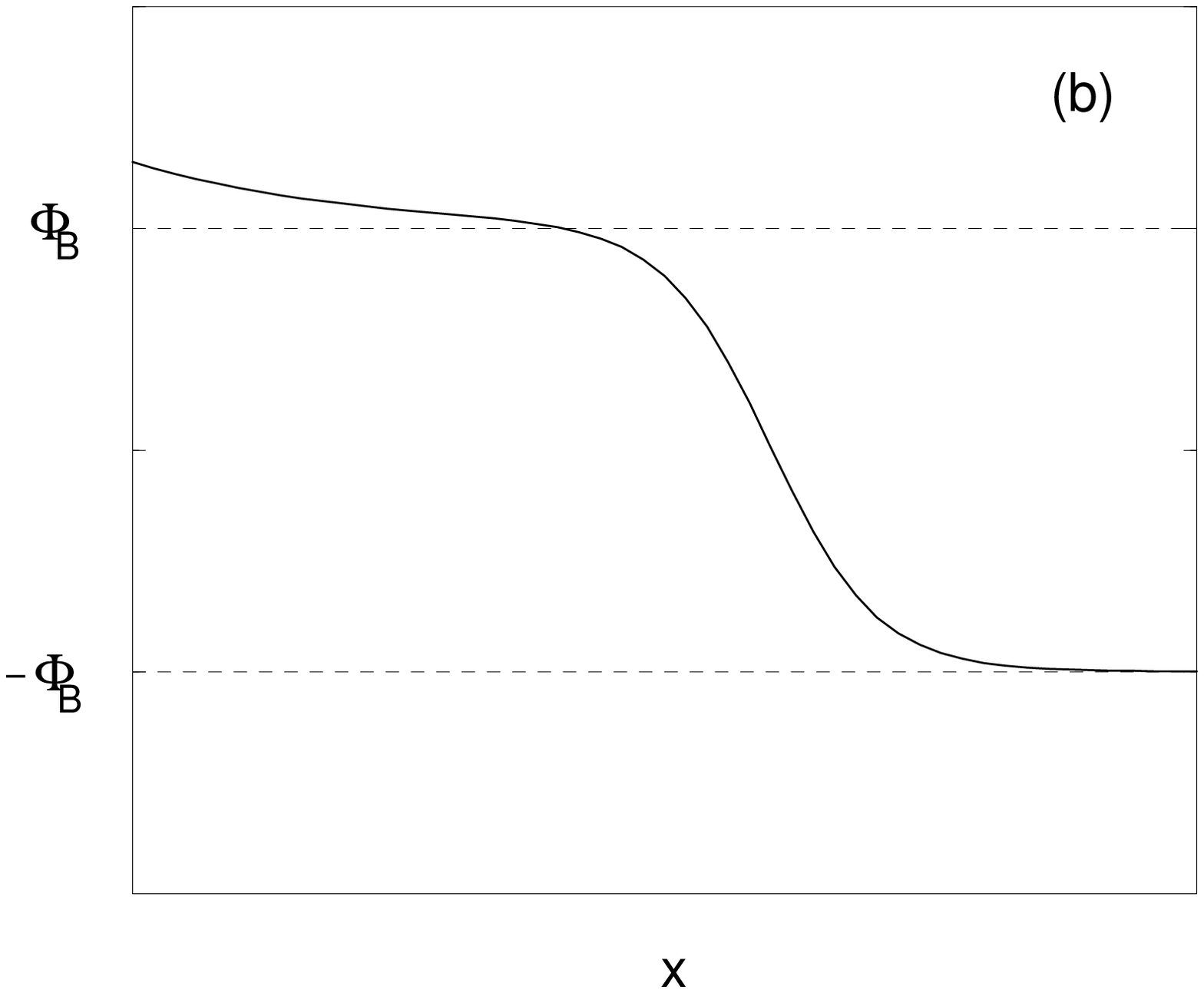}} \vspace{0.3cm}
\caption{Schematic order parameter profiles for $h < 0$ and $h_s >
0$, $(a)$ at large negative $r$ and (b) close to the bulk critical
line. In the limit $h \to 0^-$, the width of the surface layer of
solution $(b)$ diverges, leading to a wet state.}
\end{figure}
\begin{equation}
\label{integrated1} {{d \phi} \over dx} = -|\phi + \phi_B|
\sqrt{{1 \over 2}(\phi - \phi_B)^2 - {h \over \phi_B}}
\end{equation}
where the $(-)$ sign in the right hand side is taken since for the
choice of the bulk and surface fields in this problem the order
parameter is expected to decrease with $x$.

In order to locate the wetting point $WT$ we take the limit $h \to
0^-$ in Eq. (\ref{integrated1})
\begin{equation}
\label{integrated2} {{d \phi} \over dx} = -{1 \over{\sqrt
2}}|\phi^2 - \phi_B^2|
\end{equation}
In order to evaluate the surface free energy, $F_s$, associated
with the local surface order we note that the free energy density
of the bulk state is given by $C$. Thus $F_s = F - CL$. Using
(\ref{freenergy}), (\ref{bulkOP}) and (\ref{C}) one obtains
\begin{equation}
\label{surfaceenergy}
 F_s = F -CL=\int_0^L dx(\phi^{\prime})^2
 -h_s \phi_s
 \end{equation}
or
\begin{equation}
\label{surfaceenergy1}
 F_s =\int_{\phi_s}^{-\phi_B} d \phi {{d \phi} \over dx}  -h_s \phi_s
\end{equation}
%
%
%
%
%
%
%
%
We proceed by evaluating the order parameter profile obtained from
Eq. (\ref{integrated2}). Two distinct types of profiles are found:
$\phi_1(x)$ for negative and large $r$ and $\phi_2(x)$ for
negative and small $r$, close to the bulk critical point. These
profiles are schematically given in Figure 2. For large $r$, the
surface field does not affect the local surface order in a
substantial way and thus the surface order parameter $\phi_{s1}$
remains close to the bulk value, satisfying $-\phi_B < \phi_{s1} <
\phi_B$. Integrating (\ref{integrated2}) one finds the surface
free energy for this type of  solution
\begin{equation}
\label{surfaceFE1} F_{s1}={{\sqrt 2} \over 3} (\phi_B^3
+\phi_{s1}^3)
\end{equation}
where the surface order parameter, $\phi_{s1}$, is determined by
the boundary equation
\begin{equation}
\label{surfaceOP1} h_s = {1 \over {\sqrt 2}} (\phi_B^2 -
\phi_{s1}^2)
\end{equation}
On the other hand for small $r$ the local order is highly
susceptible to the local ordering field and one obtains an order
parameter profile $\phi_2(x)$ which at the surface is considerably
different from the bulk value, satisfying $\phi_{s2} > \phi_B$.
Integrating (\ref{integrated2}) for this solution one obtains the
surface free energy
\begin{equation}
\label{surfaceFE2} F_{s2}= {\sqrt 2}\phi_B^3 -{{\sqrt 2} \over 3}
\phi_{s2}^3
\end{equation}
with the surface order parameter satisfying
\begin{equation}
\label{surfaceOP2} h_s = {1 \over {\sqrt 2}} (\phi_{s2}^2 -
\phi_B^2)
\end{equation}
At the wetting transition the two types of solutions have the same
free energy. To find the transition point we define $y_1=\phi_{s1}
/ \phi_B$, $y_2=\phi_{s2} / \phi_B$. At the coexistence point one
has
\begin{equation}
\label{wettingeq} y_1^3 + y_2^3 =2
\end{equation}
In addition, the boundary condition equations (\ref{surfaceOP1})
and (\ref{surfaceOP2}) impose the following relation between $y_1$
and $y_2$ :
\begin{equation}
\label{wettingbc} y_1^2 + y_2^2 =2
\end{equation}
To solve Eqs. (\ref{wettingeq}) and (\ref{wettingbc}) we use the
substitutions $y_1^2=1-u$ and $y_2^2=1+u$ where $u= h_s {\sqrt 2}/
\phi_B^2$. Equation (\ref{wettingeq}) is then readily solved
yielding $u^2 = {\sqrt {12}} - 3$. The wetting transition thus
takes place at
\begin{equation}
\label{wettinpt} h_s= -{r \over {\sqrt 2}} {\sqrt{ {\sqrt {12}}
-3}}
\end{equation}

To locate the critical prewetting $CP$ point we consider the
boundary equation (\ref{boundary}). Combining it with the
expression for the order parameter derivative (\ref{integrated1})
it may be written as
\begin{equation}
\label{G} G(\phi_s) \equiv \phi_s^4 +2r\phi_s^2 -4h\phi_s
+(\phi_B^4 - 2h\phi_B - 2 h_s^2) =0
\end{equation}
This equation determines the surface order parameter $\phi_s$ for
given $r,h$ and $h_s$. At the critical prewetting point the two
solutions of this equation, which correspond to the two coexisting
states on the prewetting line become identical. The conditions for
this to take place are $\partial G/ {\partial \phi_s}=
\partial ^2 G / {\partial \phi_s^2}=0$. These equations
together with (\ref{G}) yield the critical prewetting point
\begin{eqnarray}
\label{CP} r &=& -\sqrt{2 \over 3} h_s \nonumber \\ h &=& -2({2
\over {27}})^{3/4} h_s^{3/2}
\end{eqnarray}

It is of interest to explore the general validity of this global
phase diagram by considering Landau free energies with different
non-linear terms. To this end we studied the phase diagram of a
Landau model with piecewise parabolic potential
\begin{equation}
\label{parabolicFE}
 F=\int_0^L dx(-h \phi + f(\phi)
+ {\textstyle{1\over2}} (\phi^{\prime})^2 ) -h_s \phi_s
\end{equation}
where
\begin{equation}
\label{potential} f(\phi) = \left\{ \begin{array}{ll} {1 \over 2}
a^2 (\phi - \phi_0)^2 & \mbox {$\phi \ge 0$} \\ {1 \over 2} a^2
(\phi + \phi_0)^2 & \mbox {$\phi < 0$}
\end{array}
\right.
\end{equation}
and the parameters $a$ and $\phi_0$ are dependent on $r$.
The analysis presented above for the Nakanishi-Fisher model may be
extended to study the phase diagram of the Landau free energy
(\ref{parabolicFE}). It is found that the $(r,h)$ phase diagram of
the two models exhibit the same qualitative features. The wetting
transition takes place at
\begin{equation}
\label{wettingPara} h_s = {1 \over 2} a \phi_0
\end{equation}
while the critical prewetting point is found to be located at
\begin{eqnarray}
\label{CPPara} h_s &=& 2 a \phi_0 \nonumber \\ h &=& -a^2 \phi_0
\end{eqnarray}
%
%
     %

\section{Lifshitz Point Model}

In this section we study the surface phase diagram corresponding
to a model of a Lifshitz point in the presence of a surface
ordering field. We consider the Landau free energy
\begin{eqnarray}
\label{freenergy1} &F& = - h_s \phi_s \nonumber \\ &+&\int_0^L
dx(-h \phi + {\textstyle{1\over2}}r \phi^2 + {\textstyle{1\over4}}
\phi^4 + {\textstyle{1\over2}}v (\phi^{\prime})^2 +
{\textstyle{1\over2}} (\phi^{\prime \prime})^2)
\end{eqnarray}
with $v=0$. This model for $v>0$ will be considered in the next
section. The Euler-Lagrange equation corresponding to this free
energy is
\begin{equation}
\label{euler1} \phi^{\prime\prime\prime\prime} -v
\phi^{\prime\prime} -h + r\phi+\phi^3=0
\end{equation}
with the boundary equations at 
\begin{equation}
\label{bcs} -v \phi_s^\prime + \phi_s^{\prime\prime\prime}-h_s=0\
, \ \ \ \ \phi_s^{\prime\prime}=0 \ .
\end{equation}

We have integrated numerically  Eq. (\ref{euler1}) with the
boundary conditions (\ref{bcs}) for $v=0$, corresponding to the case
of a Lifshitz point. We find that unlike the ferromagnetic case,
the model does $\it not$ exhibit a wet phase. It rather exhibits
two distinct phases (a prewet state, $PW$, and a state, $S$, with
surface enhanced order) both characterized by surface states with
a finite width. Here, in analogy with the phase diagram of the 
Nakanishi-Fisher model we keep referring to the low temperature 
state as the prewet phase, although the phase diagram does not
exhibit a wet state at all.
The resulting $(r,h)$ phase diagram for $h_s = 0.1$ is
given in Fig. 3. The diagram displays a first order transition
line separating the two surface phases $PW$ and $S$. The line terminates
at a surface critical point $SC$. This line intersects the $h=0$ axis
at the point $PS$.
\begin{figure}
\epsfxsize=7.0cm \centerline{\epsffile{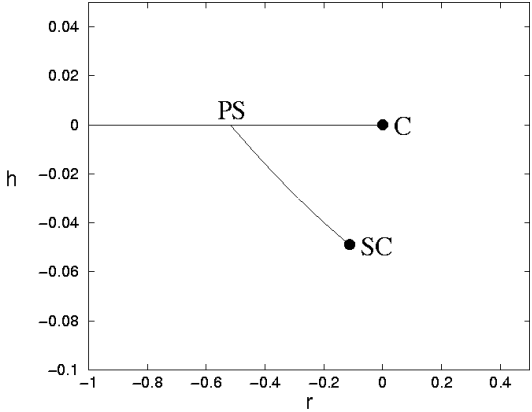}}
\vspace{0.3cm} \caption{The $(r,h)$ phase diagram corresponding to
the Lifshitz point model for $h_s = 0.1$. The phase diagram
displays two non-wet states: a prewet state, $PW$, and a state
with surface enhanced order, $S$. The two phases are separated
by a first order line. This line intersects the $h=0$ axis at the
point $PS$. It terminates for finite $h$ at a surface critical point
$SC$.}
\end{figure}

Representative order parameter profiles in the two non-wet phases
are given in Fig. 4 for small bulk field $h$. At low temperatures,
$r \ll -1$, the surface field introduces only a weak local order
near the surface, and the order parameter decays monotonically to
its bulk value as one moves away from the wall. This is very
similar to the low temperature phase of the ferromagnetic case.
However at higher temperatures, just below the bulk critical
point, the order parameter becomes highly susceptible to the local
surface field, the surface order parameter is much larger than the
bulk value, and it decays in a non-monotonic way to the bulk value
away from the wall. The width of the surface layer remains finite
even in the limit $h \to 0^-$. This is in a sharp contrast with
the ferromagnetic case where the width diverges in this limit,
leading to a wet state.

\begin{figure}
\epsfxsize=7.0cm \centerline{\epsffile{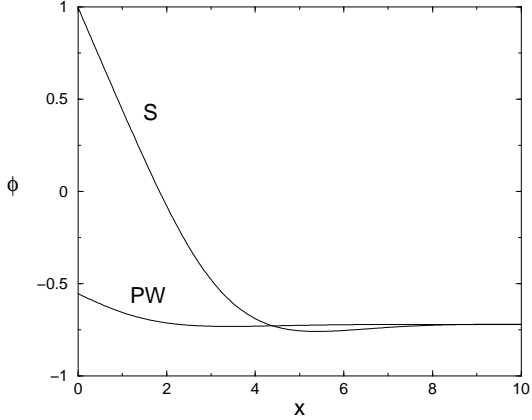}}
\vspace{0.3cm} \caption{Characteristic profiles of the order
parameter of the two non-wet states $PW$ and $S$. The profiles are
given at a point on the coexistence line of the two states, close to
the point $PS$ of Fig. 3 where $h \to 0^-$,
$r=-0.52008, h_s=0.1, h=-10^{-6}$.}
\end{figure}
%
%
%
%
                  %

\section{Extended Ferromagnetic Model}

In this section we consider the extended ferromagnetic model
(\ref{freenergy1}) with $v = 1$. We restrict this study to
negative bulk fields in the limit $h \to 0^-$ and evaluate the
$(r,h_s)$ phase diagram.
\begin{figure}
\epsfxsize=7.0cm \centerline{\epsffile{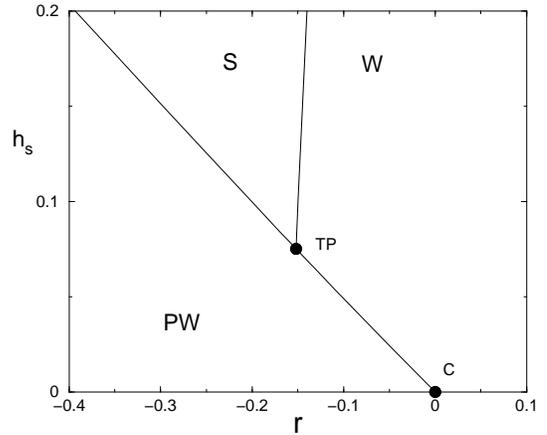}}
\vspace{0.3cm} \caption{The $(r, h_s)$ phase diagram of the
extended ferromagnetic model in the limit $h \to 0^-$.
It displays three distinct states: a wet state, $W$, and
two non-wet states $PW$ and $S$. The state $S$ is characterized
by surface enhanced order (see Fig. 6). The phases are
separated by three first order line which intersect at a 
triple point $TP$. The bulk critical point is denoted by $C$.}
\end{figure}

For small surface fields the model is found to yield similar
surface phenomena as the model of Nakanishi and Fisher. It
exhibits a wet phase, $W$, at temperatures below the bulk critical point
and a prewet state, $PW$ at low temperatures. The two states are
separated by a first order wetting transition.

However the phase diagram becomes rather different for large
$h_s$. Here, in addition to the wet state existing at high
temperatures, one finds $\it two$ distinct surface phases at lower
temperatures: a phase with a surface enhanced order, $S$, and
a prewet phase, $PW$. The two phases are
separated from each other by a first order
transition. The resulting $(r,h_s)$ phase diagram is given in Fig.
5. We also display some characteristic order parameter
profiles of the three phases. Fig. 6 gives the profiles at a point
on the $PW-S$ coexistence line, while the profiles at a point
on the $PW-W$ line are given in Fig. 7.  

\begin{figure}
\epsfxsize=7.0cm \centerline{\epsffile{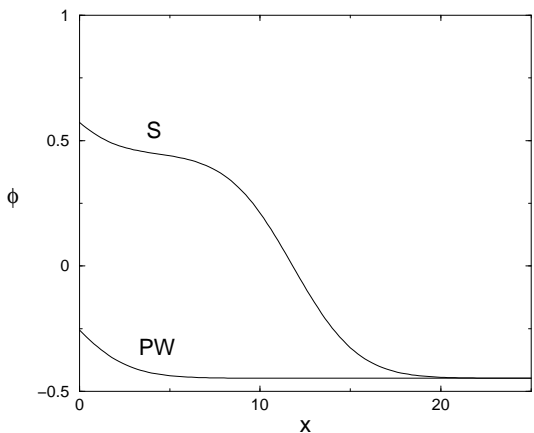}}
\vspace{0.3cm} \caption{Characteristic profiles
of the order parameter at a point along the $PW-S$ 
coexistence line for $h \to 0^-$. The profiles are given for
$r=-0.2, h_s=0.09967$ and $h=-10^{-10}$. It is evident that while 
in the $S$ state the surface order is highly susceptible to the
surface field, the surface order in the $PW$ state is affected only
mildly by this field.}
\end{figure}
%
%
%
%
%
%
%

\section{Summary}

Using a general Landau model, we have studied the surface phase
diagrams that result from the effect of a wall bounding a
semi-infinite sample which exhibits homogeneous bulk phases.  The
model is applicable to a wide variety of systems including
magnetic materials and cholesteric liquid crystals.

Choosing coefficients to simplify the model to the ferromagnetic
model studied by Nakanishi and Fisher, we obtained analytic
expressions for the temperature and bulk field co-ordinates of the
wetting transition point and the critical prewetting point as
functions of the surface field.  We also demonstrated that the
general surface phase diagram is not highly sensitive to the
precise form of the non-linear terms.
\begin{figure}
\epsfxsize=7.0cm \centerline{\epsffile{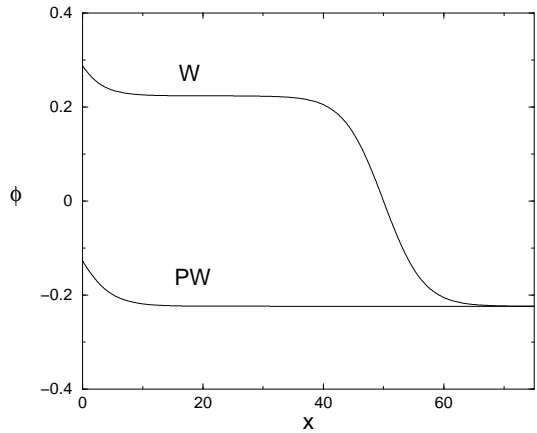}}
\vspace{0.3cm} \caption{Characteristic profiles of the
order parameter at a point along the $PW-W$ coexistence line
for $h \to 0^-$. The profiles are given for
$r=-0.05, h_s = 0.02433$ and $h=-10^{-10}$. Note that in this
figure the flat part of the $W$ state has been truncated,
as the width of the surface layer of this state diverges in
the limit of vanishing bulk field.}
\end{figure}

In contrast, we found that a system at the Lifshitz point is
similar to the case of modulated bulk phases where a wetting layer
does not form. Instead, a non-wet surface state forms which decays
non-monotonically.  The wetting transition is replaced by a
transition between the two types of non-wet states and the
critical prewetting point becomes a surface critical point.

In the extended ferromagnetic model we examined only the case of
vanishing bulk field and found that all three types of surface
layers develop if the surface field is not too weak.  As
temperature is reduced from the bulk critical point, a wetting
transition occurs, followed at even lower temperatures by a
transition from a highly deviated to a weakly deviated non-wet
layer.  Unfortunately, we are not aware of any experimental
observations of such a transition so far.  This is, however, not
surprising because it is a subtle change that has not been
expressly looked for.
\acknowledgments

We thank Michael Schick for helpful comments.
This research was supported by the Natural Sciences and
Engineering Research Council of Canada, by the Meyerhoff
Foundation, and by the National Science Foundation under
Science and Technology Center ALCOM Grant No.
DMR 89-20147.

%
%

\end{document}